\documentclass[%
 reprint,superscriptaddress,amsmath,amssymb,prb]{revtex4-1}

\usepackage{graphicx}
\usepackage{dcolumn}
\usepackage{bm}
\usepackage{hyperref}
\usepackage{physics}
\usepackage{epstopdf}
\usepackage{natbib}
\setcitestyle{square,numbers,sort&compress}
\usepackage{CJKutf8}

\begin{document}


\begin{CJK}{UTF8}{gbsn}
\title{Comparing many-body localization lengths via non-perturbative construction of local integrals of motion}
\author{Pai Peng (彭湃)}
\affiliation{Department of Electrical Engineering and Computer Science, Massachusetts Institute of Technology, Cambridge, MA 02139} \affiliation{
Research Laboratory of Electronics, Massachusetts Institute of Technology, Cambridge, Massachusetts 02139, USA
}
\author{Zeyang Li}
\affiliation{Department of Physics, Massachusetts Institute of Technology, Cambridge, MA 02139}
\affiliation{
Research Laboratory of Electronics, Massachusetts Institute of Technology, Cambridge, Massachusetts 02139, USA
}
\author{Haoxiong  Yan}
\affiliation{
Research Laboratory of Electronics, Massachusetts Institute of Technology, Cambridge, Massachusetts 02139, USA
}
\author{Ken Xuan Wei}
\affiliation{Department of Physics, Massachusetts Institute of Technology, Cambridge, MA 02139}
\affiliation{
Research Laboratory of Electronics, Massachusetts Institute of Technology, Cambridge, Massachusetts 02139, USA
}
\affiliation{IBM T.J. Watson Research Center, Yorktown Heights, NY 10598, USA}
\author{Paola Cappellaro}\email[]{pcappell@mit.edu}
\affiliation{Department of Nuclear Science and Engineering, Massachusetts Institute of Technology, Cambridge, MA 02139}
\affiliation{
Research Laboratory of Electronics, Massachusetts Institute of Technology, Cambridge, Massachusetts 02139, USA
}

\begin{abstract}
Many-body localization (MBL), characterized by the absence of thermalization and the violation of conventional thermodynamics, has elicited much interest both as a fundamental physical phenomenon and for practical applications in quantum information. 
A phenomenological model, which describes the system using a complete set of  local integrals of motion (LIOMs), provides a powerful tool to understand MBL, but can be usually only computed approximately. 
Here we explicitly compute a complete set of LIOMs with a non-perturbative approach, by maximizing the overlap between LIOMs and physical spin operators in real space. The set of LIOMs satisfies the desired exponential decay of weight of LIOMs in real-space. This LIOM construction  enables a direct mapping from the real space Hamiltonian to the phenomenological model and thus enables studying the localized Hamiltonian and the system dynamics. 
We can thus study and  compare the localization lengths extracted from the LIOM weights, their interactions, and dephasing dynamics, revealing interesting aspects of many-body localization. Our scheme is immune to accidental resonances and can be applied even at phase transition point, providing a novel tool to study the microscopic features of the phenomenological model of MBL.
\end{abstract}

\maketitle
\end{CJK}

\section{Introduction}
How a many-body quantum system thermalizes --or fails to do so-- under its own interaction is a fundamental yet elusive problem. Localization serves as a prototypical example for the absence of thermalization, first studied in the non-interacting single particle regime known as Anderson localization \cite{Anderson1958, Abrahams2010}, and then revived in the context of interacting systems (many-body localization, MBL) \cite{Abanin2018}. The existence of MBL as a phase of matter was demonstrated theoretically \cite{Basko2006, Imbrie2016a, Imbrie2016} and numerically \cite{Znidaric2008, Pal2010, Oganesyan2007, Berkelbach2010, Gornyi2005}. Recently, the MBL phase was observed in cold atoms \cite{Schreiber2015a, Choi2016, Bordia2016a, Kondov2015, Lukin2018, An2018}, trapped ions \cite{Smith2016, Roushan2017} and natural crystals using nuclear magnetic resonances \cite{Wei2018}. 
Most characteristics of MBL, such as area law entanglement \cite{Serbyn2013, Bauer2013}, Poisson level statistics \cite{Pal2010, Oganesyan2007}, logarithmic growth of entanglement \cite{Bardarson2012, Znidaric2008, Abanin2013, Huse2014a, Vosk2013, Lukin2018, Kim2014a} and power law dephasing \cite{Serbyn2014a, Serbyn2014, DeTomasi2017, Chen2017, Serbyn2017}, can be understood via a phenomenological model that expresses the Hamiltonian in terms of a complete set of local integrals of motion (LIOMs) \cite{Huse2014a, Serbyn2013}. 
However, the explicit computation of LIOMs and their interactions is a challenging task, complicated by the fact that the set of LIOMs is not unique. 
LIOMs have been calculated by the infinite-time averaging of initially local operators \cite{Chandran2015, Geraedts2017}, however, the obtained LIOMs does not have binary eigenvalues and thus cannot form complete basis. Binary-eigenvalue LIOMs can be obtained using perturbative treatment of interactions \cite{Imbrie2016a, Ros2015, Rademaker2016, Rademaker2017, You2016}, Wegner-Wilson flow renormalization \cite{Pekker2017}, minimizing the commutator with the Hamiltonian \cite{OBrien2016}. The previous methods either requires strong disorder field strength, or assumes a cutoff of LIOMs in real space, so a complete numerical study of localization lengths is missing. 

Here we design and implement a method to compute a complete set of binary LIOMs (i.e., with eigenvalues $\pm1$) in a non-perturbative way, by maximizing the overlap with physical spin operators. 
This criterion enables a recursive determination,  similar to quicksort, of the LIOMs matrix elements in the energy eigenbasis, without the need to exhaust all the eigenstate permutations, which would be prohibitive for system size $L>5$. 
We verify that in the MBL phase the LIOMs are exponentially localized in real space, and their interaction strength decays exponentially as a function of interaction range. 
LIOM localization lengths and interaction localization lengths can be extracted from the two exponential behavior respectively. Deep in the MBL phase, the two localization lengths are well characterized by the inequality derived in Ref.~\cite{Abanin2018}. 
Near the transition point, that our construction enables exploring, the interaction localization length diverges, while the LIOM localization length remains finite: this should be expected given the constraints imposed by our construction, even if it contradicts the  inequality in Ref.~\cite{Abanin2018}. 
The explicit form of the LIOMs further enables exploring the system evolution, and we show that the LIOMs display a similar dynamics to the physical spin operators, and extract the dynamical localization length from the power law dephasing process~\cite{Serbyn2014a}. 
Interestingly, we find that the dynamical localization length is much shorter than would be given by a conjectured relationship to the above two localization lengths~\cite{Serbyn2014a, Abanin2018}, suggesting that the dynamics does not only depend on the typical value of LIOMs and their interactions, but also on higher order correlations. 

\section{Algorithm}
\begin{figure}[!htbp]\centering
\includegraphics[width=0.46\textwidth]{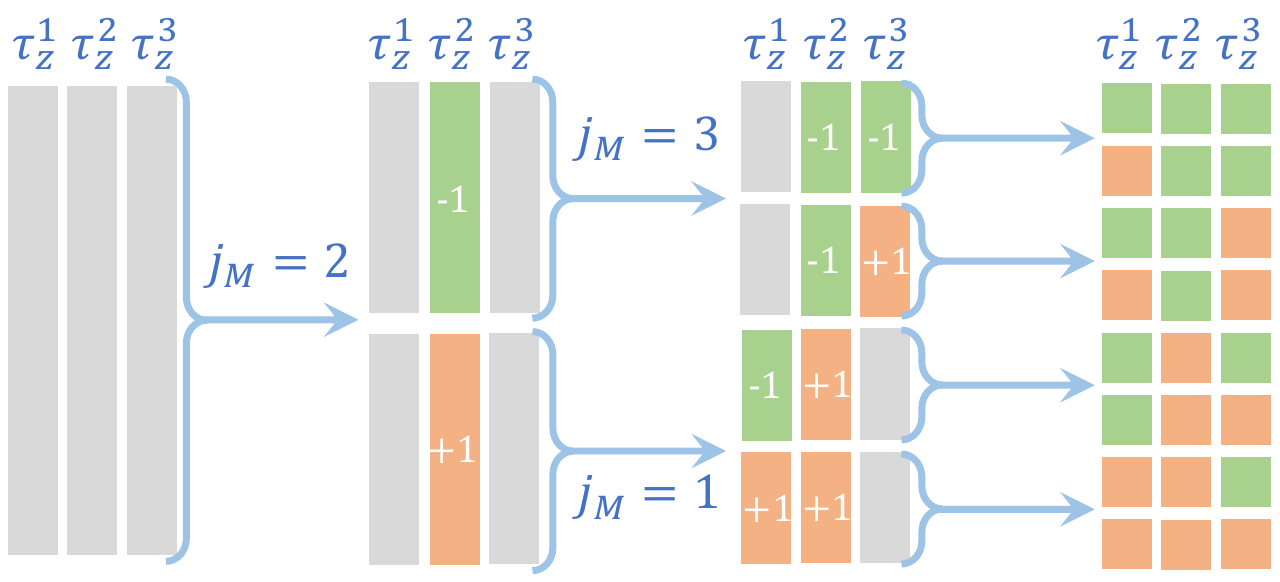}
\caption{\label{fig:schematic}
The flow diagram shows an example of the construction of a complete set of LIOMs in a system with $L=3$. Grey block represents undetermined matrix element and orange (green) block represents +1 (-1) matrix element. First diagonalize the Hamiltonian, then find $j_M$ that maximizes $\langle \tilde{\tau}_z^j \sigma_z^j\rangle$ ($j_M=2$ here). Divide the 8 eigenstates into two sectors each containing 4 states according to $\langle n|\sigma_z^2|n\rangle$ and assign $\tau_z^2=\tilde{\tau}_z^2$. For each sector, find $j_M$ within the sector, divide into two sectors each containing 2 states and assign $\tau_z^{j_M}=\tilde{\tau}_z^{j_M}$. Repeat the step one more time and then all LIOMs are determined. }
\end{figure}

To understand the construction algorithm, we first review the properties of integrals of motions in the many-body localized phase. LIOMs $\{\tau_z^j\}$ are diagonal in the Hamiltonian  eigenbasis $[H,\tau_z^j]=0$. 
A complete set of LIOMs can be related to physical spin operators by a local unitary transformation $\tau_z^j=U\sigma_z^j U^\dagger$, which implies that (i) half of the eigenvalues of $\tau_z^j$ are +1 and the other half are -1; (ii) LIOMs are  mutually independent (orthonormal) $\mathrm{Tr}(\tau_z^j \tau_z^k)/2^L=\delta_{jk}$; (iii) the weight of $\tau_z^j$ decays exponentially in real space for localized Hamiltonians. 
In particular, property (ii) requires that, for any $j$, in either +1 or -1 sector of $\tau_z^j$, half of the diagonal elements of $\tau_z^k$ are +1 and the other half are -1 for all $k\neq j$. In another word, the +1 and -1 sectors of $\tau_z^j$ are effectively two manifolds that represent two instances of a new system with $L-1$ spins, containing all sites except $j$.

With only  constraints (i-ii), there are $2^L!/L!$ different sets of IOMs among which we want to find the most local one. 
However, enumerating the $2^L!/L!$ different sets, and quantifying the localization of the related $\tau_z^j$, is numerically prohibitive. 
Therefore, instead of explicitly demanding the exponential localization, we maximize the overlap of LIOMs and physical spin operators $\mathrm{Tr}(\tau_z^j \sigma_z^j)$, which enables a systematic and efficient way to find a unique set of LIOMs, and then we verify that these LIOMs are indeed exponentially localized in the MBL phase.

Expanding the IOMs $\tau_z^j$ in the energy eigenbasis $\{\ket{n}\}$, $n=1,2,\cdots,2^L$, as $\tau_z^j=\sum_n a^j_n |n\rangle\langle n|$,  our goal is to find $a_n^j\in\pm1$ under the constrains (i-iii).  (We thus assume that we have diagonalized the Hamiltonian). 
The algorithm 
is reminiscent of quicksort (see Fig. \ref{fig:schematic}):

\begin{enumerate}
\item For all eigenstates $|n\rangle$ and spin $j$, evaluate $s_n^j=\langle n|\sigma_z^j|n\rangle$.
    \item For each $j$, sort the eigenstates according to $s_n^j$, and define  candidates $\tilde\tau_z^j=\sum_{n\in S^j_{max}}\ket{n}\bra{n}-\sum_{n\in S^j_{min}}\ket{n}\bra{n}$, where $S^j_{max(min)}$ is the set of  eigenstates giving the $2^L/2$ largest (smallest) overlaps $s_n^j$. 
\item For each $j$, compute the overlaps $\langle\tilde\tau_z^{j}\sigma_z^{j}\rangle=\sum_{n\in S^j_{max}}s_n^j-\sum_{n\in S^j_{min}}s_n^j$ and find the site $j_M$ that maximizes it. For this site, set $\tau_z^{j_M}\equiv\tilde\tau_z^{j_M}$.
\item Consider the two manifolds $\mathbb S^{j_M}_\pm$ 
corresponding to the $\pm1$ eigenstates of $\tau_z^{j_M}$. Each of these manifolds represents two instances of a new system with $L-1$ spins,  containing all sites except $j_M$. In this new system, perform the same protocol 1-3 to set another LIOM. This results in 4 sectors, each containing $2^{L-2}$ states.
    \item By repeating the previous steps  $L-2$ times we finally reduce the dimension of each sector to just 1  and all $a_n^j$ are assigned. 
\end{enumerate}

We note that our scheme does not necessarily find the most local set of $\tau_z^j$, since once the matrix elements of a LIOM are determined at a given step, the subsequent search for the rest of the LIOMs is restricted to its perpendicular complement to satisfy orthogonality (that is, we are not ensured to find a global optimum).  
Therefore, we choose to divide sectors using the most local LIOM (largest $\langle \tilde\tau_z^j \sigma_z^j\rangle$), so that this division sets the least constrains to  later divisions.
In  Fig. \ref{fig:AveH_seq} of the Appendix/SM we show that this choice indeed gives the most local results among all alternate algorthms we tried. Because we only utilize the overlaps $s_n^j=\langle n|\sigma_z^j|n\rangle$ in the computation, the scheme is immune to accidental resonances in the spectrum.

\section{Results}
\begin{figure}[!htbp]\centering
\includegraphics[width=0.48\textwidth]{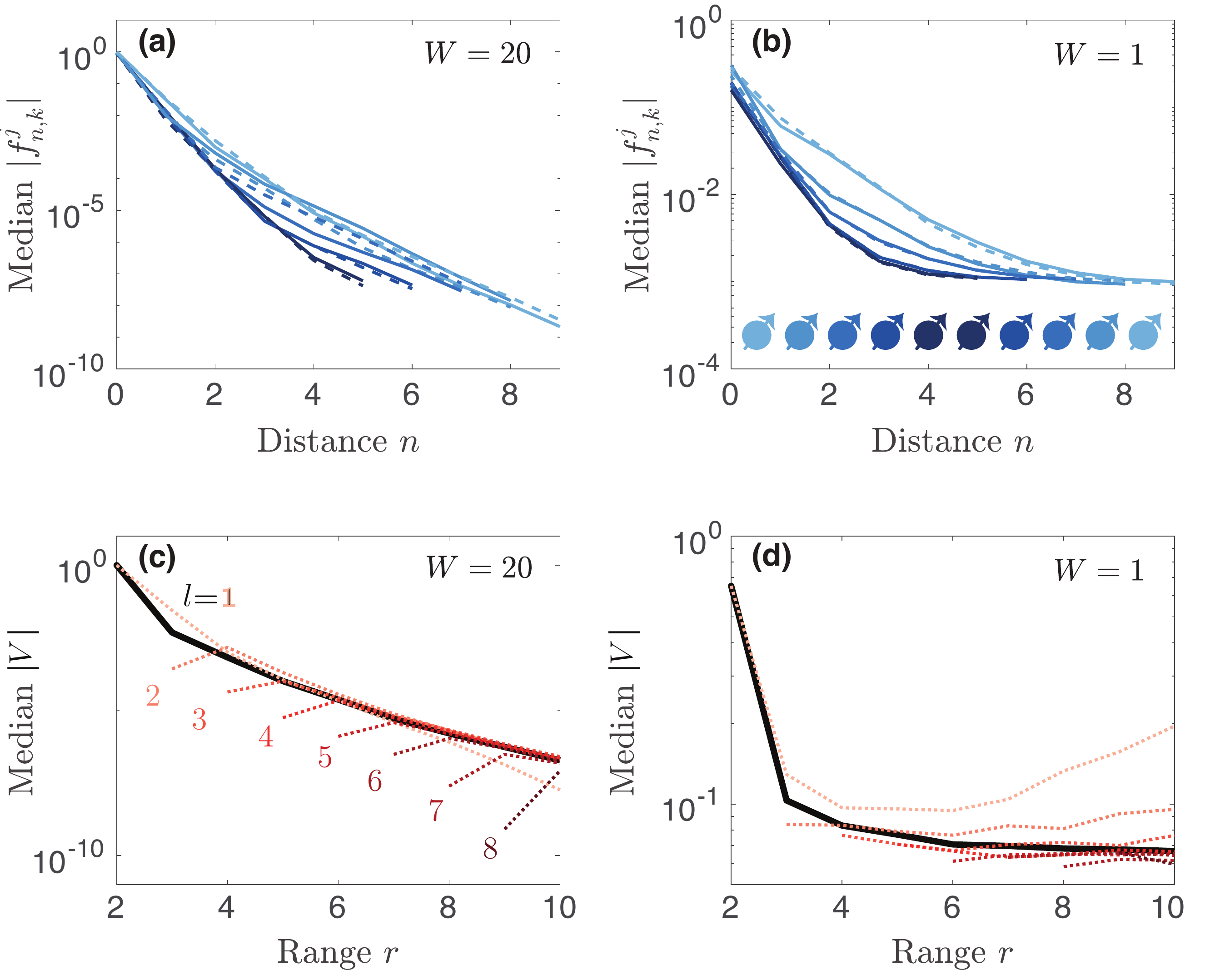}
\caption{\label{fig:TauzH}
(a-b) Median of the LIOM weights $|f_{n,k}^j|$ as a function of distance $n$ for two disorder strengths: (a) $W=20$,  deep in the MBL phase, where the median decays exponentially; and (b)  $W=1$, in the ergodic phase, where the median saturates. For each $j$, the median is taken over the index $k$ in $|f_{n,k}^j|$ as well as 20 different disorder realizations. Darker color represents LIOMs in the middle of the chain [as shown in the bottom of (b)], and left (right) half of the LIOMs are represented by dashed (blue) curves.
(c-d) Median of the interaction strength as a function of range $r$ for two disorder strengths. Dotted curves represent $l$-body interaction terms $ |V_{ij}|,  |V_{ijk}|, \cdots$ ($l=2,\dots,9$), where the median is taken over all indices $i,j,\cdots$, as well as 100 disorder realizations. The solid curve represents median of all interaction terms for a given range $V(r)$, regardless of how many LIOMs are involved. $L=10$ in all subplots.
}
\end{figure}

\subsection{Localization of operators and interactions}
To test the proposed algorithm and characterize the LIOMs that it finds we  consider a prototypical example of an MBL-supporting system, a Heisenberg spin-1/2 chain with random fields,
\begin{equation}
\label{eq:Hamiltonian}
H=\sum_{i=1}^L h_i \sigma_z^i +\sum_{i=1}^{L-1}\vec{\sigma}^i\cdot \vec{\sigma}^{i+1},
\end{equation}
where $h_i$ is uniformly distributed in $[-W,W]$. It is known~\cite{Pal2010} that in the thermodynamic limit there is a MBL phase transition at $W_c\approx 7\pm2$. Although this model conserves the total magnetization along $z$, the validity of the algorithm does not depend on this symmetry. To quantitatively check the locality of LIOMs, we decompose them into tensor products of Pauli operators
\begin{equation}
\tau_z^j=\sum_{n=0}^L \sum_k f_{n,k}^j \hat{O}^j_{n,k},
\end{equation}
where $\hat{O}^j_{n,k}$ is a tensor product of Pauli operators whose furthest non-identity Pauli matrix from $j$ is of \textit{distance} $n$, e.g. $\sigma_x^1 \otimes \sigma_x^2 \otimes \sigma_y^3 \otimes \mathbb{I}^4$ is of distance $n=2$ to $j=1$, because $\sigma_y^3$ is the furthest non-identity Pauli matrix. $k$  labels  operators with the same $n$. $f_{n,k}^j=\mathrm{Tr}(\tau_z^j \hat{O}^j_{n,k})$ is the weight of $j$-th LIOM on $\hat{O}^j_{n,k}$. 
Figures~\ref{fig:TauzH}(a) and (b) show the median of $|f_{n.k}^j|$ as a function of distance $n$. In the MBL phase, the median weight decays exponentially with distance $n$, while in the ergodic phase it saturates at large $n$.

Because the LIOMs form an orthonormal basis, the Hamiltonian can be decomposed into this basis unambiguously and efficiently:
\begin{equation}
\label{eq:Htauz}
H=\sum_i \xi_i\tau_z^i+\sum_{ij} V_{ij}\tau_z^i\tau_z^j+\sum_{ijk} V_{ijk}\tau_z^i\tau_z^j\tau_z^k+\cdots.
\end{equation}
For non-interacting models, only the  $\xi_i$ coefficients are nonzero. We can define the range $r$ of each coupling term $V_{ij\cdots}$  as the largest difference among the indices. For example, the range for 2-body interaction $V_{ij}$ is simply $r=|i-j|$, while for 3-body interactions  is $r=\mathrm{max}(|i-j|,|i-k|,|j-k|)$. 
Figures~\ref{fig:TauzH}(c) and (d) show the median interaction strength as a function of interaction range. In the MBL phase, the interaction strength decays exponentially. 
The behavior of two-body interactions $|V_{ij}|$ and three body interactions $|V_{ijk}|, \cdots$ show no significant difference~\cite{Rademaker2016, Pekker2017} and can be essentially captured by the median of all interaction terms for a given range $V(r)$. 
We considered the median instead of the mean in order to exclude rare events, i.e., instances where the disorder strength is small in a local region. 

\begin{figure}[t]\centering
\includegraphics[width=0.48\textwidth]{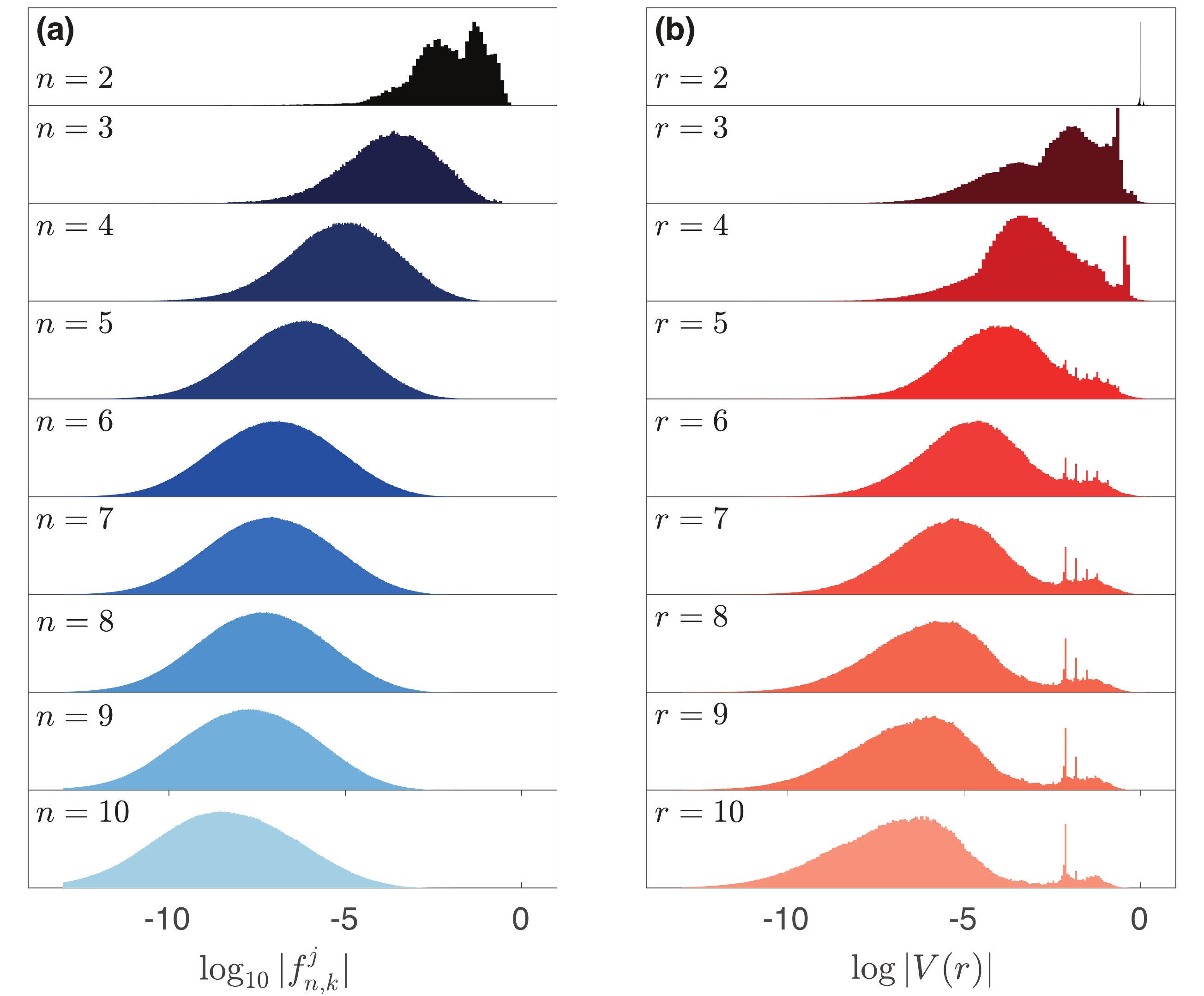}
\caption{\label{fig:PDF}
(a) Probability distribution of LIOM weights $\log_{10}|f_{n,k}^j|$. For a given $n$, samples are taken from all possible $j$ and $k$ as well as 200 disorder realizations. The distribution shows one single Gaussian peak that shifts toward smaller weights with increasing distance $n$, signaling the localization of IOMs. (b) Probability distribution of the interaction strength $\log_{10}(|V|)$. For given range $r$, samples are taken from all terms in Eq. \ref{eq:Htauz} as well as 10000 disorder realizations. Two peaks can be observed: the left peak is due to the localized cases as it shifts to smaller interaction strengths for longer range; the right peak shows the delocalized cases (rare events) as it is independent of interaction range. $L=10$ and $W=20$ for both (a) and (b). 
}
\end{figure}

\begin{figure}[!htbp]\centering
\includegraphics[width=0.49\textwidth]{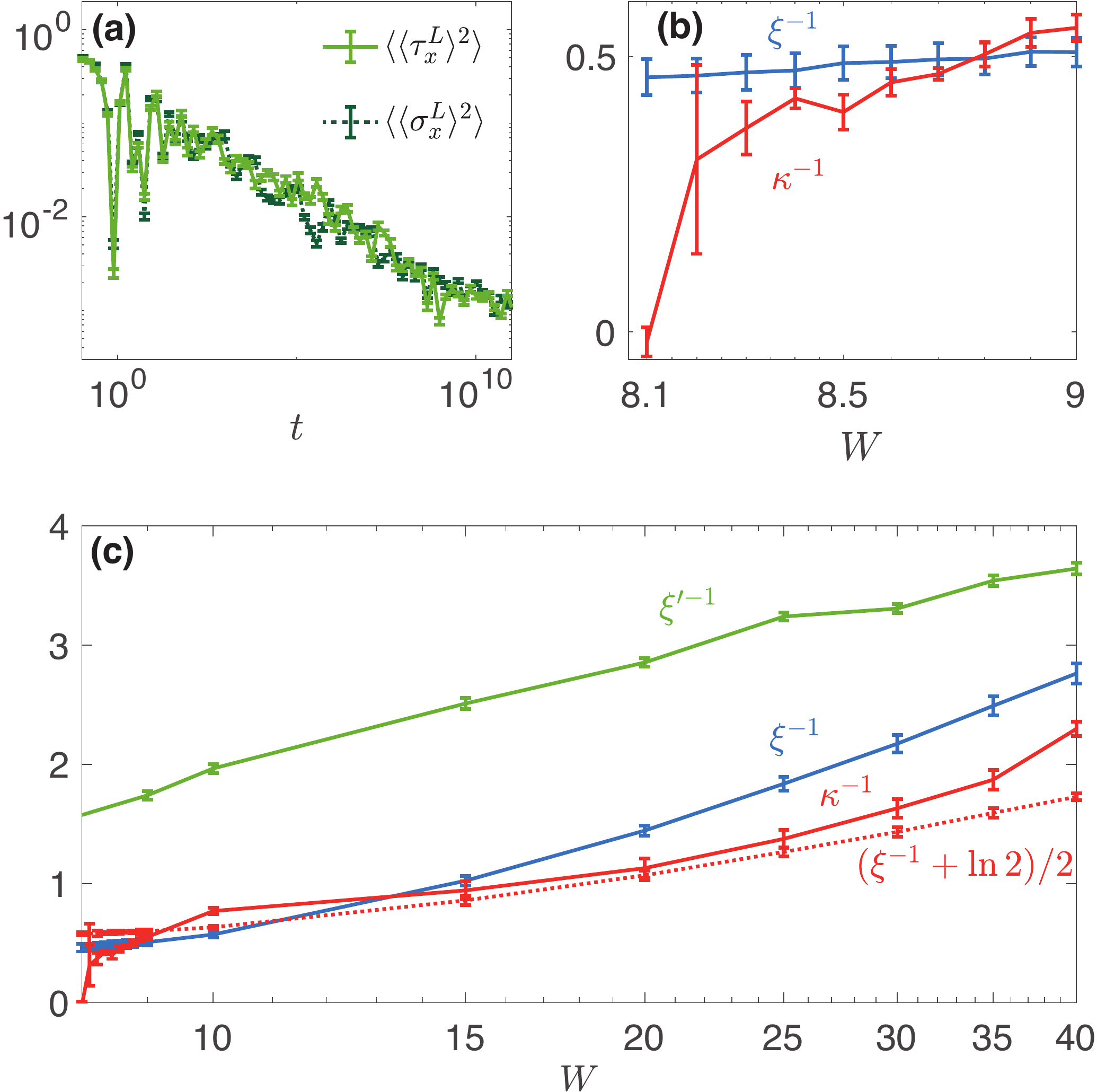}
\caption{\label{fig:ll6}
(a) Dephasing of the physical spin operator $\sigma_x^L$ (dark green, dashed curve) and LIOM $\tau_x^L$ (green, solid curve). Initial state is a product state with each spin pointing randomly in xy plane, i.e. $|\psi(0)\rangle=\otimes_{j=1}^L (|+\rangle_j +e^{i\phi}|-\rangle_j)/\sqrt{2}$, with $\phi$ randomly sampled in $[0,2\pi]$, $\sigma_z^j|+\rangle_j=|+\rangle_j$ $\sigma_z^j|-\rangle_j=-|-\rangle_j$ for red curve and $\tau_z^j|+\rangle_j=|+\rangle_j$ $\tau_z^j|-\rangle_j=-|-\rangle_j$ for blue curve. $L=10$, $W=20$. Averaging is performed over 20 different initial state and 20 disorder realizations. Error bar represents the standard deviation of all configurations.
(b) and (c) Localization length as a function of disorder strength $W$ for $L=12$. The LIOM localization lengths are extracted from $\mathrm{Tr}(\tau_z^j\sigma_z^k)\sim \exp(-|k-j|/\xi)$ with $j=1$, interaction localization lengths from $V(r)\sim\exp(-r/\kappa)$ and dynamical localization lengths from $\langle\langle\tau_x^L\rangle^2\rangle\sim t^{-\xi'\ln2}$. 
Error bar only shows the fitting error. $\xi'$ curve is extracted from the median of 50 disorder realizations and 50 initial states. $\xi$ and $\kappa$ are extracted from the median of 5000 disorder realizations. (b) is a zoom-in of (c) near the transition point.
}
\end{figure}

%
To gain more insight into the localization of IOMs and interactions and observe the occurrence of rare events, in Figure~\ref{fig:PDF} we further study the probability distribution of weight $f_{n,k}^j$ versus $n$, and the probability distribution of interaction strength $V(r)$ versus $r$ in the localized regime (strong disorder).  
The distribution of $\log(|f_{n,k}^j|)$ can be described by a single Gaussian peak, centered at smaller values of $|f_{n,k}^j|$ when the distance $n$ increases,
confirming the localization of IOMs. Instead, two peaks can be observed in the distribution of $\log(|V|)$. The left peak shifts to smaller $|V|$ with increasing $r$, while the right peak (larger $|V|$) shows no significant shift. Moreover, the area of the right peak decreases for larger $W$ and smaller $L$. Therefore, we identify the left peak as describing  localized cases, the right one as rare events. 
The exponential localization of the LIOMs and their interactions are usually the two criteria that define the LIOM. In the rare region of low disorder,  however, the two requirements cannot be satisfied simultaneously and there is no universal criteria on how to choose LIOMs in this case.
Here we require the IOM $\tau_z$ to be local by construction, so the presence of a rare region shows up only in the interaction strengths; choosing different criteria for the LIOM construction may lead to different results.

\subsection{Localization lengths}
From the explicit form of the LIOMs and their interactions, we can extract the \textit{LIOM localization length} $\xi$, via $|f_{n,k}^j|\sim \exp(-n/\xi)$, and \textit{interaction localization length} $\kappa$, via $|V(r)|\sim \exp(-r/\kappa)$~\cite{Abanin2018}. In Figure~\ref{fig:ll6} we show  $\kappa$ and $\xi$ as a function of disorder strength $W$. 
The LIOM localization length $\xi$ is extracted using the relation $\mathrm{Tr}(\tau_z^j\sigma_z^k)\sim \exp(-|k-j|/\xi)$ \cite{Chandran2015,Rademaker2016} because calculating $f_{n,k}^j$ is numerically demanding (see SM). 
The interaction localization length $\kappa$ is extracted by fitting the distribution of $\log|V|$ (as in Fig.~\ref{fig:PDF}) to two Gaussian peaks and then fitting the localized peak center to a linear function of $r$. Because our method forces $\tau_z$ to be local, $\xi$ is always finite, while $\kappa$ diverges around $W=8.1$ [Fig. \ref{fig:ll6}(b)], which agrees with the critical point $W_c=7\pm 2$ reported in Ref.~\cite{Pal2010}. 
It has been shown in \cite{Abanin2018} that the two localization lengths satisfy the inequality $\kappa^{-1}\ge (\xi^{-1}-\mathrm{ln}2)/2$. From the numerical results in Fig.~\ref{fig:ll6}(c), we find that this inequality is satisfied in the localized phase, except in the vicinity of the phase transition point.

\begin{figure}[t]\centering
\includegraphics[width=0.5\textwidth]{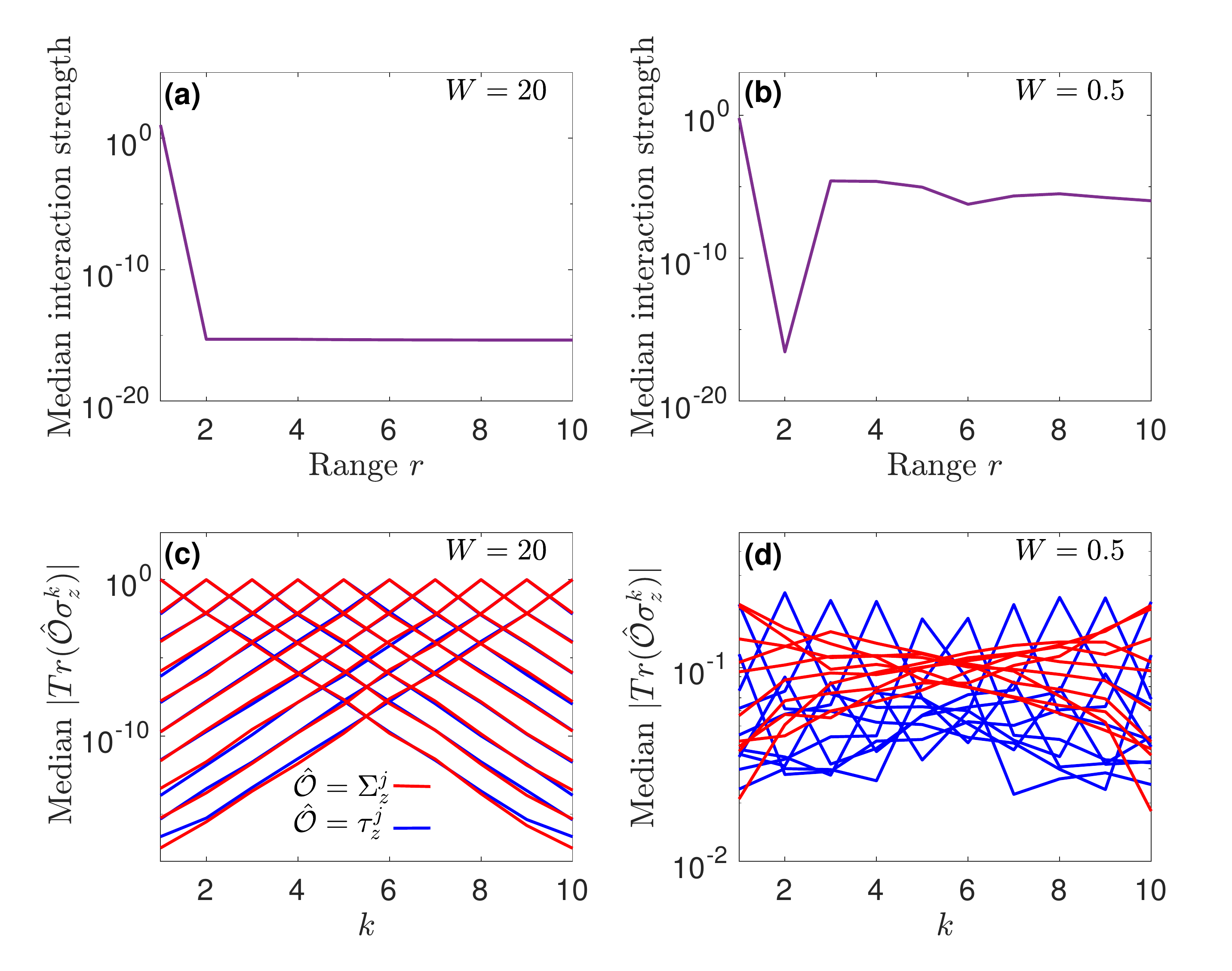}
\caption{\label{fig:lbit_int}
LIOM in non-interacting model. (a) (b) Median interaction strength in the basis of $\{ \tau_z^j\}$. $r=1$ denotes the single-particle Hamiltonian $\xi_j \tau_z^j$.
(c) (d) Median overlap between LIOMs and physical spins $\mathrm{Tr}(\hat{\mathcal{O}} \sigma_z^k)$, with $\hat{\mathcal{O}}=\Sigma_z^j$ (red) for single-particle LIOM and $\hat{\mathcal{O}}=\tau_z^j$ (blue) for LIOMs obtained using the scheme proposed in this paper. Different curves stand for different $j$.
In (a) and (c) $W=20$.  $r>1$ interaction strength is below machine precision $\sim10^{-15}$. $\Sigma_z$ and $\tau_z$ show little difference.
In (b) and (d) $W=0.5$. $\tau_z^j$ is more localized at site $j$, but the interaction among LIOMs is not zero.
L=10 and 500 disorder realizations are used in all plots.
}
\end{figure}
\subsection{Non-interacting model: tradeoff of localization}
\label{sec:noninteracting}
We can better understand why the interaction localization length $\kappa$ diverges at the critical point while the LIOM localization length $\xi$ remains finite by applying our LIOM construction to a non-interacting model 
$
H=\sum_{i=1}^L h_i \sigma_z^i +\sum_{i=1}^{L-1}\left(\sigma^i_x  \sigma^{i+1}_x+\sigma^i_y  \sigma^{i+1}_y\right)
$. Due to the lack of interactions, the system is effectively localized for arbitrarily small $W$. This Hamiltonian can be mapped to a free fermionic Hamiltonian via a Jordan-Wigner transformation \cite{Jordan1928}. The Hamiltonian can be diagonalized by single-particle IOMs $\{\Sigma_z^i\}$:
$H=\sum_i \tilde{\xi}_i \Sigma_z^i$, that is, the interaction localization length in the $\{\Sigma_z^i\}$ basis is zero. However, note that the single-particle IOMs $\{\Sigma_z^i\}$ can be highly non-local for small $W$. 
We can instead apply our algorithm to find LIOMs $\{ \tau_z^j\}$ for this model as done for the interacting Hamiltonian and compare $\{\Sigma_z^j\}$ and $\{ \tau_z^j\}$ (see Fig. \ref{fig:lbit_int}). For large disorder strength, $W=20$, the Hamiltonian is practically interaction-free even in the $\tau_z^j$ basis, and indeed the LIOMs $\tau_z^j$ approach the IOMs,  $\tau_z^j\approx\Sigma_z^j$. 
The trade off between the two interaction strength $\kappa$ and $\xi$ becomes evident for small disorder, $W=0.5$, where $\tau_z^j\neq\Sigma_z^j$. In this regime, the single-particle IOMs $\Sigma_z^j$ are delocalized, $\xi\gg1$, but the Hamiltonian still has no interactions, $\kappa=0$. Instead, the LIOMs obtained by our construction, $\{ \tau_z^j\}$, are localized but they give rise to long-range interactions in the Hamiltonian, $\kappa\gg1$. 
For interacting models, it is difficult to obtain  IOMs that minimize the interactions in a non-perturbative way. Still, we expect that if one were indeed able to find such a set of IOMs, there would be a similar tradeoff between how local they are (small $\xi$) versus how local the interactions are (small $\kappa$) outside the well-localized phase. Our choice of criterion for constructing LIOMs not only allows a simple and efficient algorithm; by keeping the operators local even when crossing the localization transition, the $\tau_z^j$ are always well-defined and can be used to explore properties of the system, such as its dynamics, around the localization-delocalization transition point. 

\subsection{Dephasing Dynamics}
Since  physical spin and LIOM operators are related by a local unitary transformation, they are expected to exhibit a similar dynamics [Fig.~\ref{fig:ll6}(a)]. In particular, the higher order interaction terms in Eq.~(\ref{eq:Htauz}) induce dephasing of the transverse operators by creating an effective magnetic field $H_\mathrm{eff}$ at the location of spin $j$ due to all the other spins. 
The  dephasing of the expectation values $\langle \tau_x(t)\rangle$ and $\langle \sigma_x(t)\rangle $ is closely related to the logarithmic light cone in the MBL phase \cite{Serbyn2014a}. It was previously shown that $\langle \langle \sigma_x(t)\rangle ^2\rangle \approx \langle \langle \tau_x(t)\rangle ^2\rangle\propto t^{-\alpha}$, where we took the average of the  expectation values  over random initial states and disorder realizations. 
For an initial state given by a product state with each individual spin pointing randomly in the xy plane, $\alpha=2\xi' \ln2$ for bulk spins and $\alpha=\xi' \ln2$ for boundary spins, where $\xi'$ is a localization length different from $\xi$ and $\kappa$ \cite{Serbyn2014a}. This length $\xi'$, that we name  \textit{dynamical localization length}, describes the strength of the contribution to the effective magnetic field felt by spin $j$ due to spins at distance $l$: $H_\mathrm{eff}^l\sim\exp(-l/\xi')$ (see SM). 
By assuming exponentially decaying interactions, $|V(r)|=\exp(-r/\kappa)$, it was conjectured that $\xi'^{-1}\ge\kappa^{-1}+(\ln2)/2$~\cite{Abanin2018}. We find instead a much larger dephasing rate [Fig.~\ref{fig:ll6}(c)]. To investigate whether this is due solely to our LIOMs construction which does not explicitly enforces an exponentially decaying interaction strength, we artificially generate an Hamiltonian satisfying $|V(r)|\propto\exp(-r/\kappa)$ (see SM). Still, although we indeed find a power law decay, this is even  
faster than what we observe in Fig.~\ref{fig:ll6}(a). We conjecture  that the  dephasing process cannot be simply described by a mean interaction strength (the model used to justify the relationship to $\kappa$), and higher order correlations may play an important role.

\section{Conclusion and Outlook}
We provide a novel method to efficiently compute the LIOMs for MBL systems by maximizing the overlap between LIOMs and physical spin operators. The method is non-perturbative and thus immune to resonances in the spectrum, and can be applied at the phase transition point. The only quantity we use in computing the LIOMs and their interactions is the expectation value of physical spin operators on energy eigenstates $\langle n|\sigma_z^j|n\rangle$. Although we use exact diagonalization here, our scheme is compatible with renormalization group methods and matrix product state representations \cite{Khemani2016a,You2016}, which can potentially be applied to much larger system and beyond one dimension. We show the power of the constructed LIOMs by extracting  the  localization length of the LIOMs and the Hamiltonian interactions from their respective exponential decays. We also show that in the MBL phase, the LIOMs and physical spin operators exhibit similar dephasing dynamics, even if it cannot be simply explained by the typical weights of LIOMs and typical interaction strengths.


\appendix

\section{Comparison of LIOMs and physical spin operators} 
\label{app:LiomSpins}
\begin{figure}[!htbp]\centering
\includegraphics[width=0.4\textwidth]{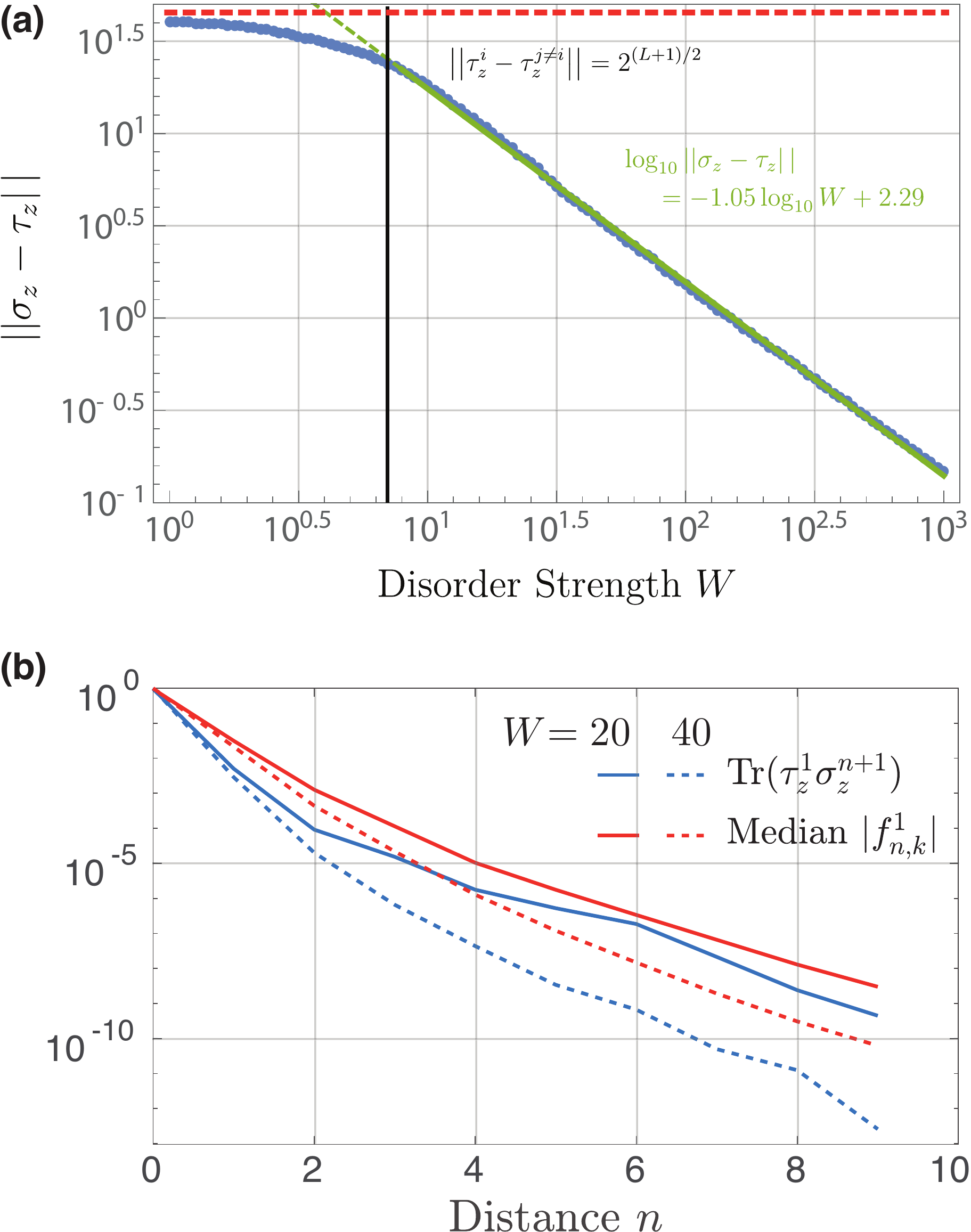}
\caption{Comparison of LIOMs and physical spins operators. (a) Blue dots show the Frobenius norm of the difference between same-site physical spin-1/2 Pauli matrices and local integral of motion, $||\sigma_z^j-\tau_z^j||$ as a function of disorder strength in a size $L=10$ system. 
For disorder $W>W_c\sim7$ (vertical black line), the norm scales as $1/W$ (green line). The red dashed line shows the norm of difference between two LIOM at different sites for comparison. (b) Weight of the first LIOM $f_{n,k}^1$ (blue curves) and overlap of the first LIOM with physical spins $\sum_j \mathrm{Tr}(\tau_z^1 \sigma_z^{1+n})/L$ (red curves) as a function of distance $n$ for $W=20$ (a) and $W=40$ (b). $L=10$ and the median is taken over $k$ and 100 disorder realizations.}
\label{fig:tautaudistance}
\end{figure}

In the main text we defined the overlap $f_{n,k}^j$ as a quantifier of the locality of the LIOMs $\tau_z^j$. Another metric that characterizes the LIOMs as a function of disorder strength is the distance of each $\tau_z^j$ from the corresponding physical spin-1/2 Pauli operator $\sigma_z^j$. Indeed, 
the larger the  disorder, the more local are the LIOMs, and therefore the closer to the corresponding Pauli operators. We use the Frobenius norm of the matrix difference between the two operators at the same site [see Fig.~\ref{fig:tautaudistance}(a)] to quantify the operator distance. At small disorder strength, the LIOM and physical spin operators are almost perpendicular,
\begin{equation}
\left.||\sigma_z^j-\tau_z^j||\right._{W\to0}\sim\sqrt{\left.||\sigma_z^j||^2\right.+\left.||\tau_z^j||^2\right.}=2^{(L+1)/2}.
\end{equation}
As the disorder strength increases, the distance decreases, as expected. At strong disorder strength $W>W_c\sim7$, we find that the distance decreases as $1/W$, indicating that the system is in the MBL phase. This result shows that the Frobenius norm distance (or equivalently the trace norm) can be taken as good proxy for the overlap $f_{n,k}^j$.

In the main text we state that the LIOM localization length can be extracted from $\mathrm{Tr}(\tau_z^j \sigma_z^k)$. To confirm this quantitatively, in Fig. \ref{fig:tautaudistance}(b) we compare the weight of first LIOM $f_{n,k}^1$ and $\text{Tr}(\tau_z^1 \sigma_z^{1+n})$. Both of them show exponential decay with $n$ and the slopes (decay rates) are similar for $n\ge2$. In numerics, calculating $f_{n,k}^1$ is demanding because it is defined in the real space (see SM), while calculating $\text{Tr}(\tau_z^1 \sigma_z^{1+n})$ can be done in the energy eigenbasis since the expectation value of $\sigma_z^j$ on every eigenstate is already obtained during the construction process. Therefore we use $\text{Tr}(\tau_z^j \sigma_z^{1+n})$ with $n\ge2$ to extract the LIOM localization length $\kappa$ in the main text (Figure~\ref{fig:ll6}).

\section{Distribution of interaction strengths and rare regions}

\begin{figure}[!htbp]\centering
\includegraphics[width=0.48\textwidth]{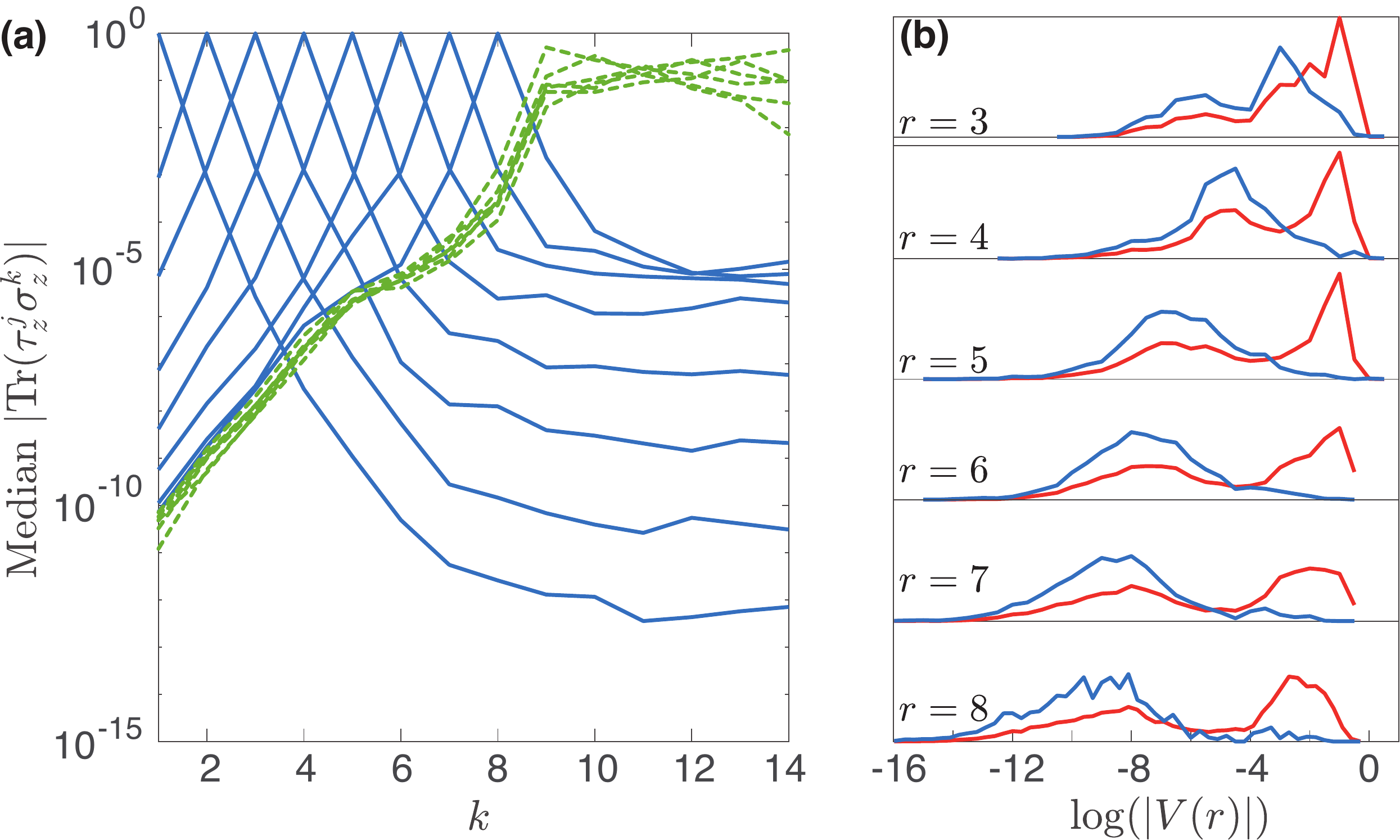}
\caption{\label{fig:PDFJ_rare}
Median $|\mathrm{Tr}(\tau_z^j \sigma_z^k)|$ (a) and normalized probability distribution of $\log|V(r)|$ (b) for a chain of size $L=14$ with disordered field only on site 1 to 8 ($i_e=8$). In (a), green dashed (blue solid) curves represent the LIOMs in the disorder-free (disordered) region. In (b), blue curves show only the coupling terms within the disordered region and red curves show the distribution of all coupling terms. $W=50$ and 100 random realizations are used in both (a) and (b).
}
\end{figure}

In the main text we linked the occurrence of rare events in the distribution of interaction strengths to rare regions of the disordered field. 
We can verify this conjecture by taking a closer look at one particular disorder realization that contains a low-disorder rare region (see Fig. \ref{fig:RareCase} in the SM). 
To further confirm the connection between a rare region and the rare event peak in the interaction strength distribution,  we study a Heisenberg spin chain with disorder field only on part of the chain, i.e. $h_i\in[-W,W]$ in the disordered region $i\le i_e$ and $h_i=0$ in the disorder-free region $i> i_e$. The LIOMs in the disordered region are localized, while the LIOMs in the disorder-free region are delocalized with an exponential tail extending into the disordered region [Fig. \ref{fig:PDFJ_rare}(a)]. 
Due to the existence of the disorder-free region, the probability distribution of $\log|V(r)|$ shows a large delocalized peak [blue curve in Fig. \ref{fig:PDFJ_rare}(b)], which is absent when considering only interactions inside the disordered region.
We can further analyze how the occurrence of rare events changes with the system size (see Figure \ref{fig:PDFJ} of the SM).
We find that for a given interaction range, the area of the delocalized peak gets larger for longer chain, because the frequency of having a local low-disorder region is higher for larger $L$. 



\section{Dephasing with Artificial Hamiltonian}
\begin{figure}[tb]\centering
\includegraphics[width=0.47\textwidth]{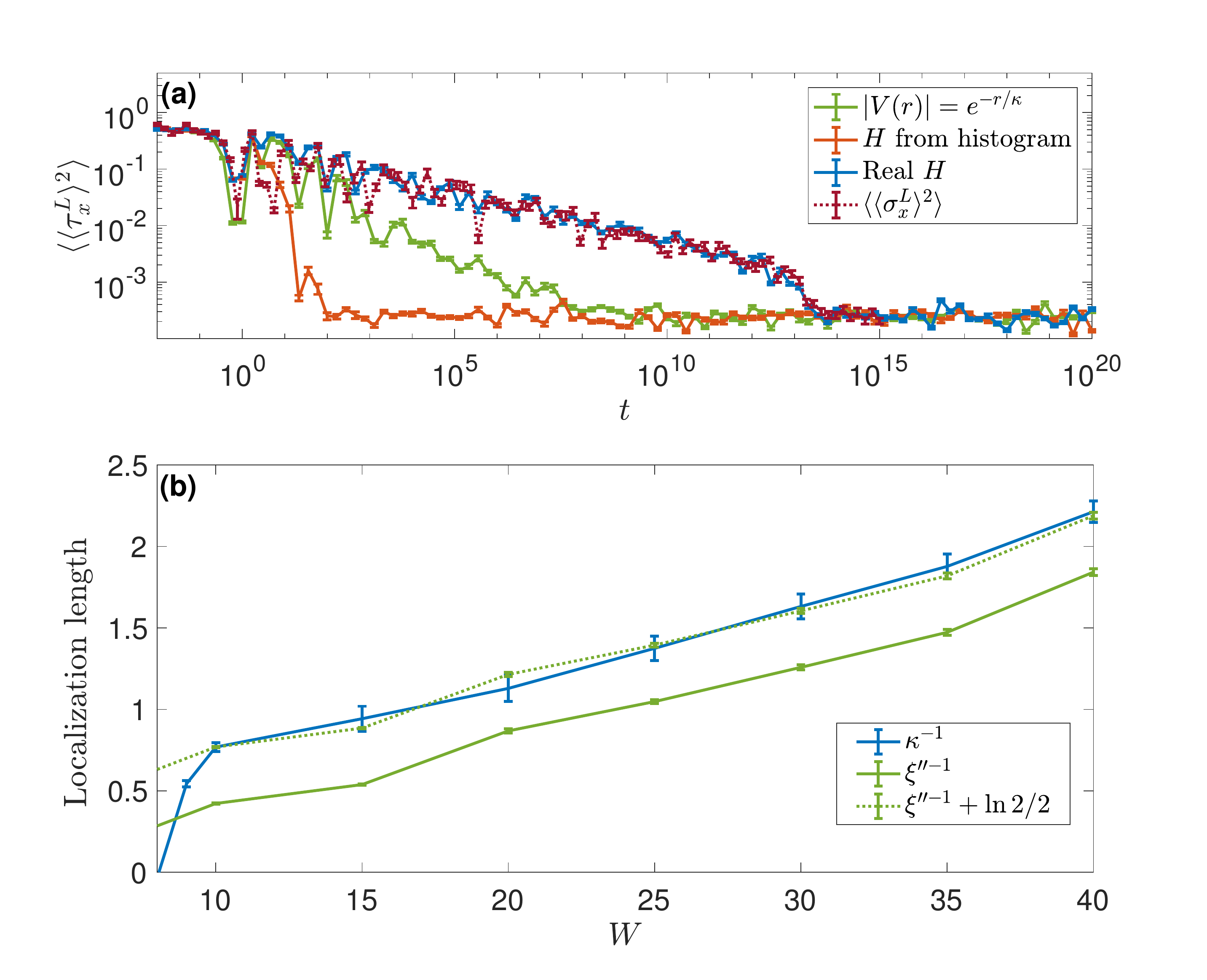}
\caption{\label{fig:dephase}
(a) $\langle\langle\tau_x^L\rangle^2\rangle$ under real (green) and two artificial Hamiltonian (blue and red). The dark green, dashed curve shows $\langle\langle\sigma_x^L\rangle^2\rangle$ under the real Hamiltonian. $L=12$, $W=40$, so that the delocalized cases is negligible. All $\langle\langle\tau_x^L\rangle^2\rangle$ are averaged over 20 random initial state in xy plane and 20 disorder realizations. $\langle\langle\sigma_x^L\rangle^2\rangle$ is averaged over 10 random initial state in xy plane and 10 disorder realizations.
(b) The red curve shows $\kappa^{-1}$ as in Fig. \ref{fig:ll6} of the main text. Green solid curve represent the dephasing localization length $\xi''$ extracted from artificial Hamiltonian with $|V(r)|=e^{-(r/\kappa)}$. Green dashed curve shows $\xi''+\ln2/2$ which overlaps with $\kappa^{-1}$ within the error bars.
}
\end{figure}
It has been conjectured that the dephasing rate of $\langle\tau_x\rangle$ (and $\langle\sigma_x\rangle$) can be related to the interaction localization length via a simple, mean-field model. Using our LIOM construction, we found instead surprising results as shown in  Figure \ref{fig:ll6}. Here we want to (i) verify whether assuming an exponentially decaying interaction strength does indeed yield the relationship between localization lengths presented in Ref.\cite{Abanin2018}; and (ii) determine whether the Hamiltonian approximation   with a simpler, exponentially decaying interaction strength is enough to capture the exact dephasing dynamics.

To answer these questions,  we consider two artificially generated Hamiltonian: (1)  $|V(r)|=\exp(-r/\kappa)$, with  each interaction term  randomly assigned a plus or minus sign,  
and (2) $|V(r)|$ randomly sampled from the simulated probability distribution (see Fig. \ref{fig:PDF}(b) in the main text for an example) with a random sign.

The first Hamiltonian exactly satisfies the hypothesis under which the relation between interaction localization length and dynamical localization length was derived in  Ref. \cite{Abanin2018}. Therefore we fit the power law dephasing obtained under this Hamiltonian [see Fig.~\ref{fig:dephase}(a)] and extract the dynamical localization length $\xi''$ as done in Fig.~\ref{fig:ll6} of the main text. We find that $\kappa^{-1}\approx\xi''^{-1}+\ln2/2$ [Fig.~\ref{fig:dephase}(b)], which gives a more stringent  relation than the bound $\kappa^{-1}\ge\xi''^{-1}-\ln2/2$ given in Ref. \cite{Abanin2018}. 
We can provide a heuristic argument for the relation between $\xi''$ and $\kappa$, under the assumption $|V(r)|=\exp(-r/\kappa)$. As described in Ref. \cite{Serbyn2014a}, the dephasing can be understood as arising from an effective magnetic field at site $j$ due to all other spins $\tau_z$.  
Starting from the phenomenological model in Eq.~(\ref{eq:Htauz}), the effective magnetic field at site $j$ is
\begin{equation}
H_j=\mathrm{Tr} (\tau_z^j H)=\xi_j +H_j^1 +H_j^2 +\cdots,
\end{equation}
where $H_j^l$ denotes the magnetic field created by spins within the distance $l$ from spin $j$. For example, the first term is given by 
\begin{equation}
    H_j^1\!=\!V_{j,j+1} \tau_z^{j+1}+V_{j-1,j} \tau_z^{j-1}+V_{j-1,j,j+1}\tau_z^{j-1}\tau_z^{j+1}.
\end{equation}
Similarly, $H_j^l$ contains interactions of range $l+1,l+2,\cdots,2l+1$. As the interaction strength decays as $|V(r)|=\exp(-r/\kappa)$ and the number of terms grows as $\sim 2^r$,  the Frobenius norm of $H_j^l$ is estimated as 
\begin{equation}
    ||H_j^l||=\left[\sum_{r=l+1}^{2l+1} 2^r e^{-2r/\kappa}\right]^{1/2}\approx \frac{\left(2e^{-2/\kappa}\right)^{(l+1)/2}}{1-2e^{-2/\kappa}}.
\end{equation}
In the last term we  assumed that $l\gg1$ and the system is deep in the MBL phase so that $2\exp(-2/\kappa)<1$. We thus find that $H_j^l$ also exhibits an exponential decay $H_j^l\propto\exp(-l/\xi'')$, with $\xi''^{-1}=\kappa^{-1}+\ln2/2$, yielding the dephasing~\cite{Serbyn2014a}   $\langle\langle\tau_x^L(t)\rangle^2\rangle\sim t^{-\xi''\ln2}$ as shown in   Fig. \ref{fig:dephase}. 

While we confirm that the dephasing under the approximated Hamiltonian satisfying $|V(r)|=\exp(-r/\kappa)$ follows the predicted relation to $\kappa$, we still find that dephasing under the ``real'' Hamiltonian is different. 
The physical spin and LIOMs under the real Hamiltonian in Eq.~\ref{eq:Hamiltonian} show similar dephasing as expected.
Under either artificial Hamiltonians,  however, $\langle\langle\tau_x^L\rangle^2\rangle$ dephases much faster than under the real Hamiltonian, suggesting that the dephasing dynamics cannot be fully captured by the interaction  localization length $\kappa$ or even the probability distribution of $|V(r)|$ [Fig.~\ref{fig:dephase}(a)]. For instance,  in the real system for a given disorder realization the interaction terms may have some correlation, which gives rise to a slower dephasing, but this is not captured by the probability distribution.

\bibliographystyle{apsrev4-1}
\bibliography{library}
\clearpage 
\section*{Supplementary material}

\subsection{Rare events in the distribution of interaction strengths and rare regions}
\begin{figure}[b!]\centering
\includegraphics[width=0.26\textwidth]{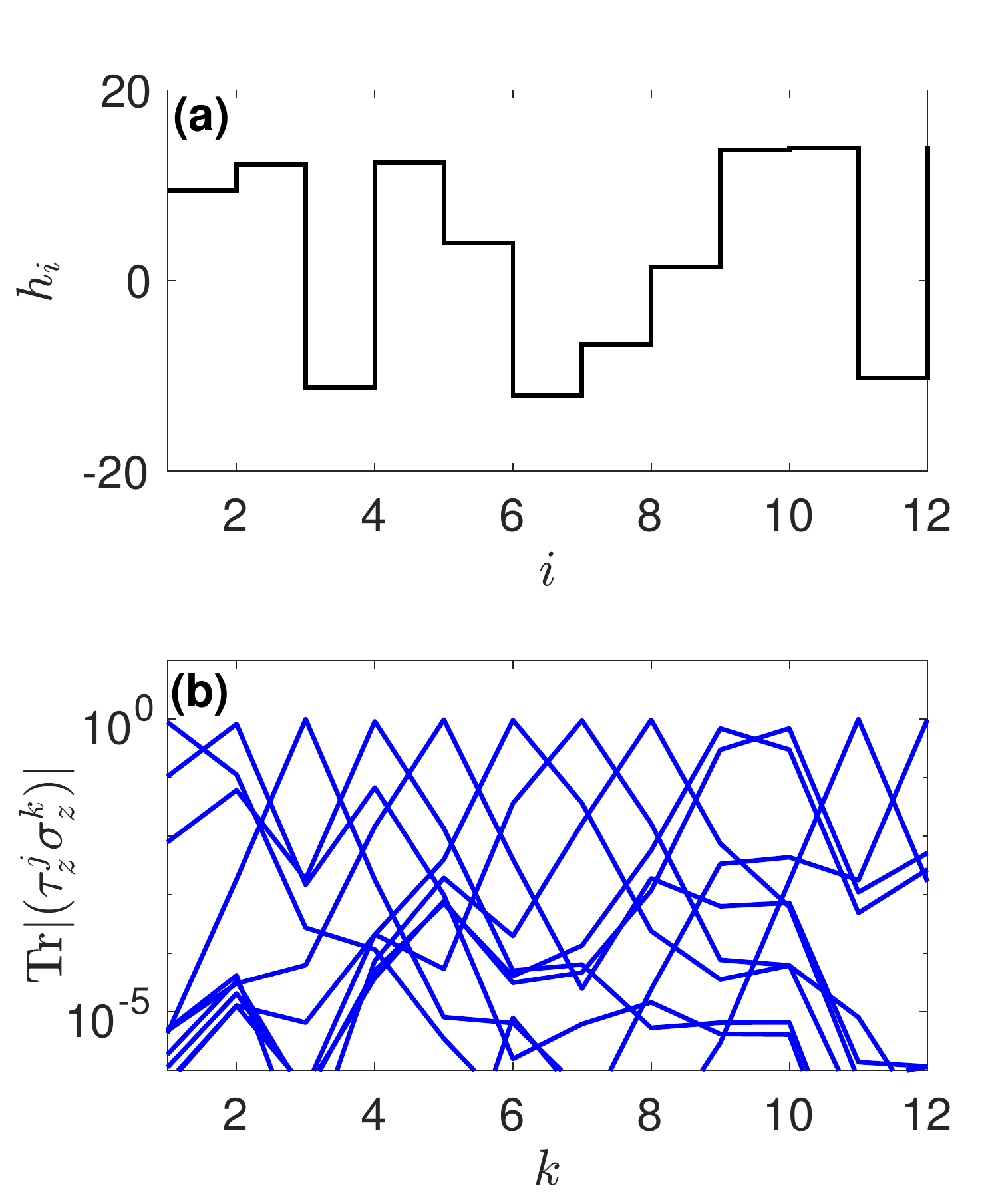}
\includegraphics[width=0.21\textwidth]{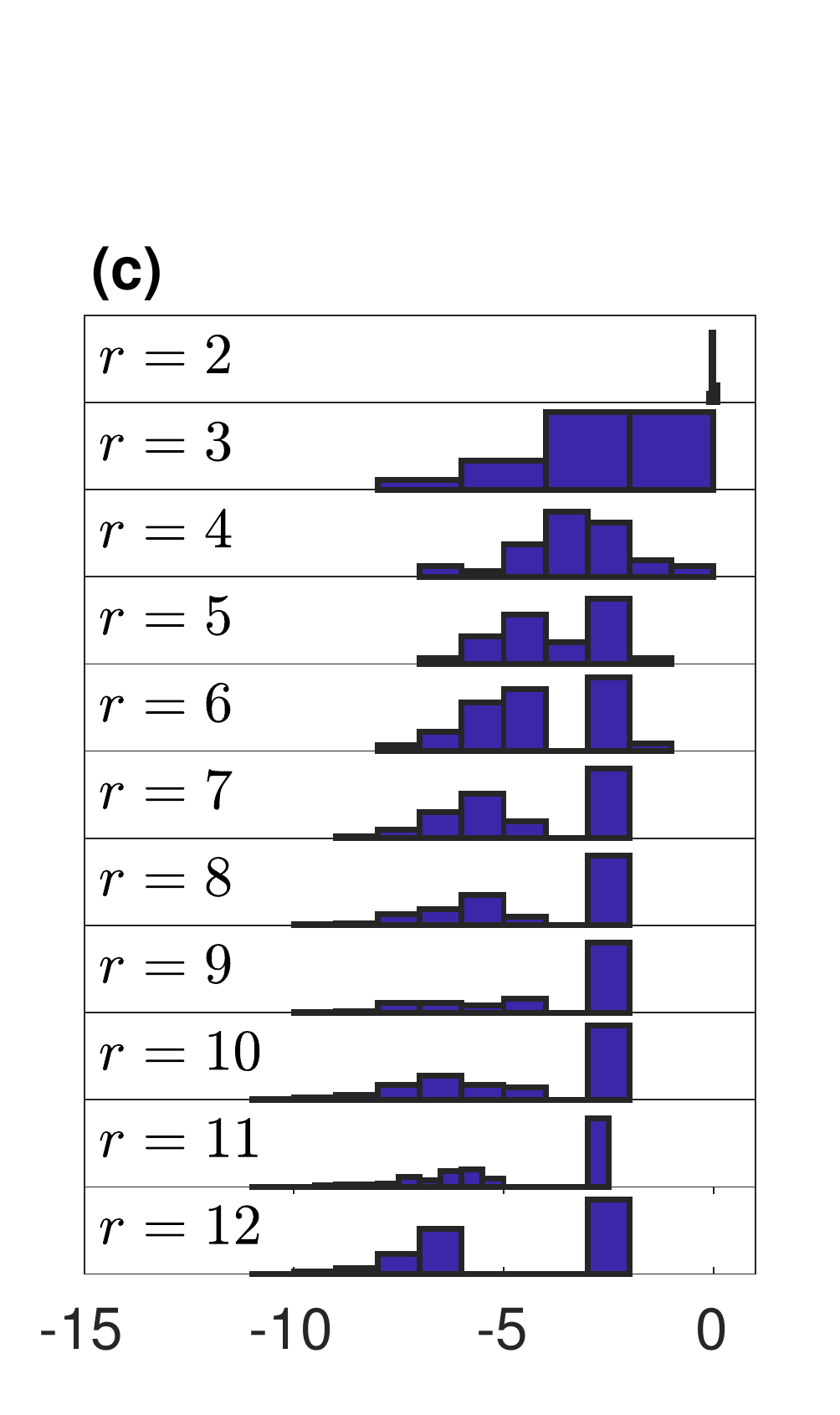}
\caption{\label{fig:RareCase}
\textbf{Results from one disorder realization that contains a rare region.} (a) Disorder field profile, where $h_9\approx h_{10}$ forms a low-disorder region. (b)
Overlap between the LIOMs and physical spins. The rare region at site 9 and 10 leads to local mix of $\tau_z^9$ and $\tau_z^{10}$. (c) Probability distribution of the logarithmic interaction strength $\mathrm{log}|V(r)|$, in which an evident peak at large $|V|$ can be observed. L=12, W=15.}
\end{figure}

\begin{figure}[b!]\centering
\includegraphics[width=0.45\textwidth]{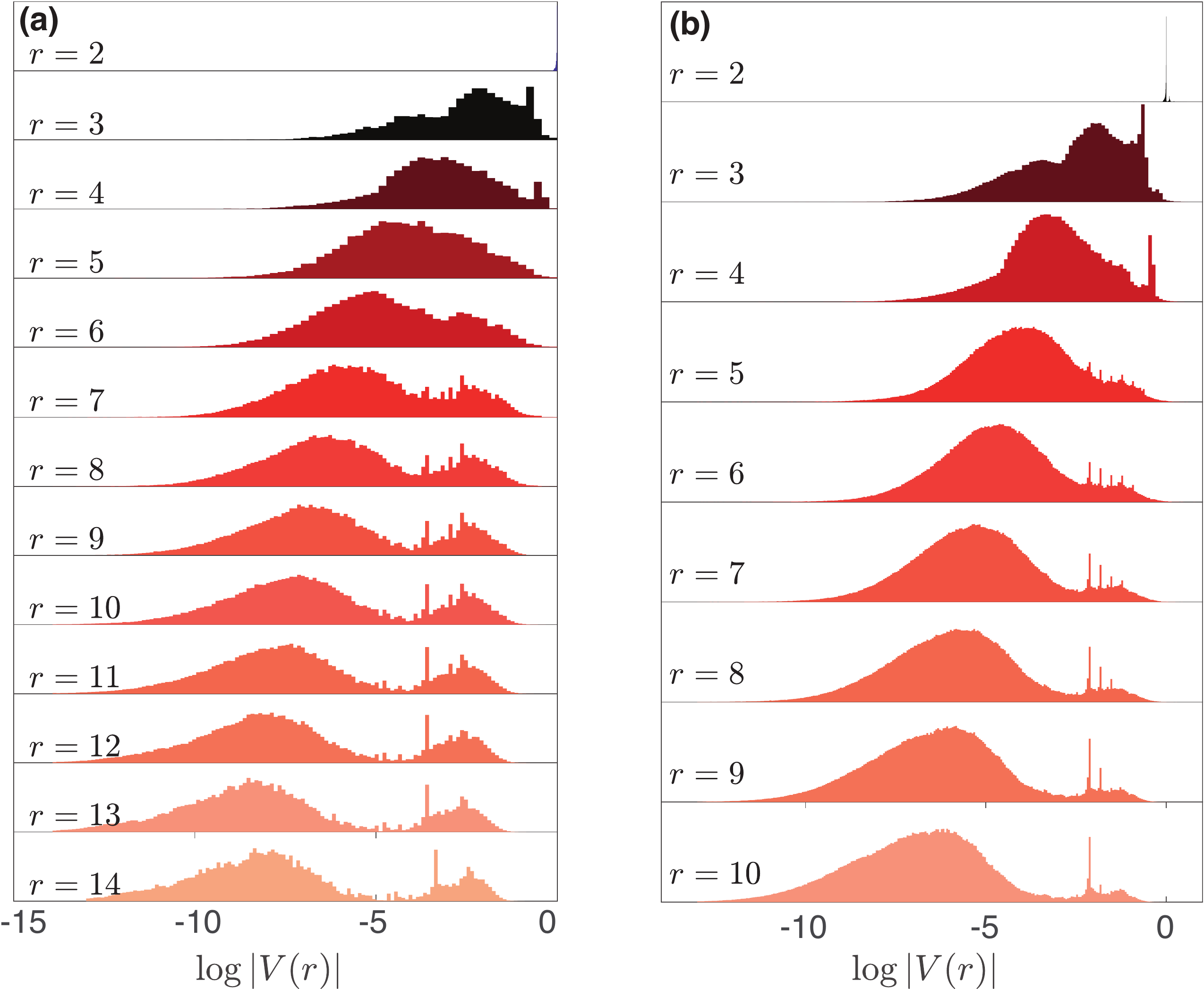}
\caption{\label{fig:PDFJ}
Probability distribution of logarithmic interaction strength $\log|V(r)|$ for $L=14$ (a) and $L=10$ (b). 400 disorder realizations are sampled for $L=14$ and 1000 for $L=10$. Two peaks can be observed: the left one shows the localized cases because it shifts to smaller $|V(r)|$ side for longer range; the right one shows the delocalized cases because it is independent of interaction range. $W=20$ in both (a) and (b).
}
\end{figure}

To verify that the peak at larger $|V(r)|$ corresponds to the rare regions, we take a closer look at one particular disorder realization that contains a low-disorder rare region (Fig. \ref{fig:RareCase}). The rare region leads to a local mix of two LIOMs, and a peak at large interaction strengths, $|V|$, arises in the interaction  distribution. 
In addition to study this link more generally, as done in the Appendix, we can also analyze the behavior as a function of the chain length. Figure \ref{fig:PDFJ} shows the probability distribution of $\log|V(r)|$ for $L=14$ and $L=10$.
For a given interaction range, the area of the delocalized peak gets larger for longer chain, because the frequency of having a local low-disorder region is higher for larger $L$. There are multiple resonances on the delocalized peak, which are not yet understood.



\subsection{Stability} 
\begin{figure}[b]\centering
\includegraphics[width=0.47\textwidth]{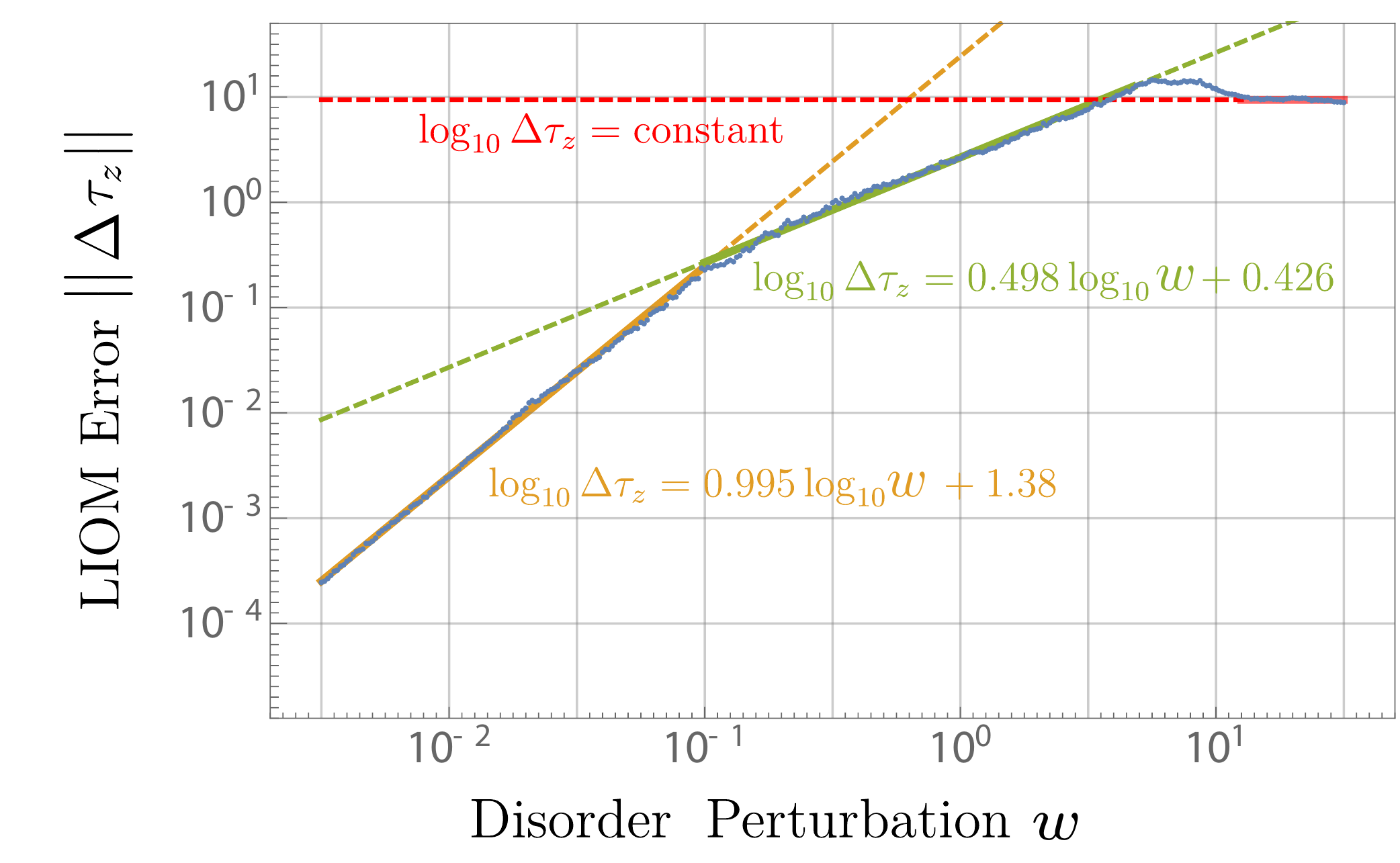}
\caption{\textbf{Stability analysis.} Median of $\log_{10}\mathrm{norm}||\Delta\tau_z||$ as a function of extra disorder strength $w$ based on original system disorder $W=20$ at $L=10$. Median is taken over 100 system disorder realizations and $L$ LIOMs for each realization. The perturbation disorder field configuration is fixed, only the scale factor $w$ varied. The fitting result for small perturbation ($w<0.01$, yellow line), has a slope close to $1$, indicating linear growth of distance. For medium extra disorder ($0.01<w<20$, green line),  the distance grows proportional to square root of the strength. For extra disorder larger than original disorder ($w>100\gg w=20$, red line), the distance saturates to a constant. There also exists a competing region of which disorder dominates.}
\label{fig:Stability}
\end{figure}

It is known that the MBL phase, contrary to integrable systems, is robust against small perturbations of the Hamiltonian; it is thus desirable that the LIOMs display the same robustness. 
To show the stability of our construction, i.e. that the LIOMs $\tau_z^j$ do not change dramatically  under small perturbations in the Hamiltonian, we generate an additional disorder field on top of the original Hamiltonian in Eq.~(\ref{eq:Hamiltonian}).  We fix the additional disorder field configuration and scale it by $w$. We quantify the deviation of the new LIOM $\tau_z^{j,w}$ from the original $\tau_z^j$ using the Frobenius norm of the matrix difference: 
\begin{align}
||\Delta\tau_z^j||=\sqrt{\mathrm{Tr}(\Delta\tau_z^{j\dagger} \Delta\tau_z^j)},
\end{align}
with $\Delta\tau_z^j=\tau_z^{j,w}-\tau_z^j$, and plot the  median of this distance as a function of $w$ in Fig. \ref{fig:Stability}. 
At first the distance grows linearly as the perturbation increases, before slowing down to a 
square root growth for intermediate perturbation, and finally saturating to a constant when the additional disorder dominates. 
The initial linear growth is expected from a  linear expansion of the operators for small perturbation strength. The eventual saturation at $w\gtrsim W$  correspond to the limit where the two operators only overlap at the same (physical) site, thus giving the maximum distance between the two operators. 
In the intermediate region, the perturbation field does not only modify the eigenstates $|n\rangle$ but might also alter some matrix elements $a_n^j$, yielding the square root scaling.
We note that the overall small deviation $\Delta\tau$ not only demonstrates
the stability of our numerical method, but also more broadly the robustness of LIOMs in the MBL phase.

\subsection{Comparison to alternative algorithms to find the LIOM set}
\begin{figure}[t]\centering
\includegraphics[width=0.49\textwidth]{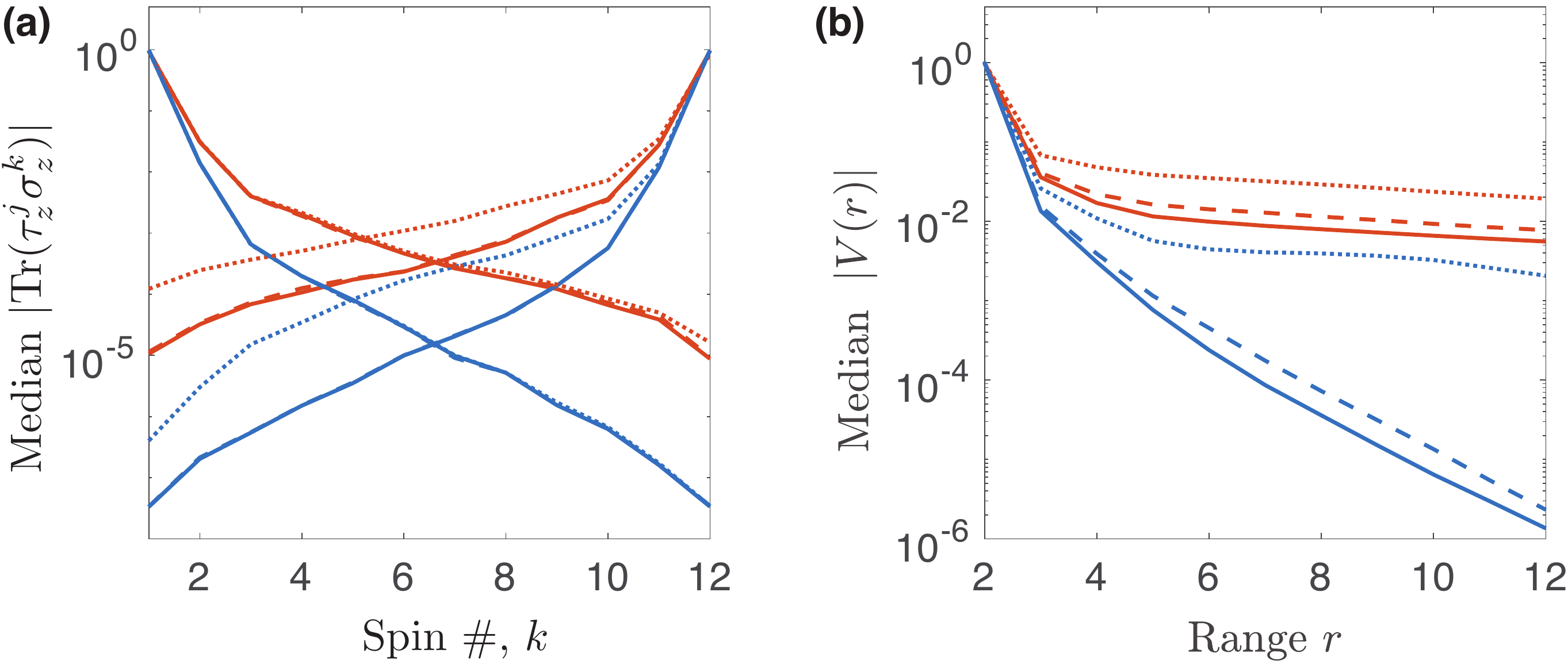}
\caption{\label{fig:AveH_seq}
\textbf{Localization of LIOMs and interactions for three schemes}: \textit{scheme 1} (solid curve), \textit{scheme 2} (dotted curve) and \textit{scheme 3} (dashed curve); blue curves are for a disorder strength $W=15$ and red curves for $W=10$. (a) Median overlap between LIOMs and physical spins $\mathrm{Tr}(\tau_z^j \sigma_z^k)$, for  $j=1,12$ as a function of spin number $k$. (b) Median interaction strength, defined in the same way as the solid curve in Fig. \ref{fig:TauzH}(c) and (d) of the main text.
For both plots we took  $L=12$ and the median was taken over 1000 realizations. 
}
\end{figure}
We compare our proposed algorithm to two similar algorithms that follow however different sector division schemes: 

\textit{Scheme 1}: recursively divide the sector according to the $j_M$ that maximizes $\langle \tilde{\tau}_z^j\sigma_z^j\rangle$ within the sector. This is the scheme used in the main text.

\textit{Scheme 2}: divide starting from the leftmost spin (smallest $j$), i.e. divide the $2^L$ states into $\mathbb{S}_+$ and $\mathbb{S}_-$ by sorting $\langle n|\sigma_z^1|n\rangle$, and then further divide $\mathbb{S}_\pm$  each into two sectors by sorting $\langle n|\sigma_z^2|n\rangle$, etc.

\textit{Scheme 3}: First, compute and sort $\langle \tilde{\tau}_z^j\sigma_z^j\rangle$ to get a sequence $j_1, j_2, \cdots j_L$ such that $\langle \tilde{\tau}_z^{j_1}\sigma_z^{j_1}\rangle>\langle \tilde{\tau}_z^{j_2}\sigma_z^{j_2}\rangle>\cdots>\langle \tilde{\tau}_z^{j_L}\sigma_z^{j_L}\rangle$. Then divide according to this sequence, i.e. divide the $2^L$ states into $\mathbb{S}_+$ and $\mathbb{S}_-$ by sorting $\langle n|\sigma_z^{j_1}|n\rangle$, and then further divide $\mathbb{S}_\pm$ each into two sectors by sorting $\langle n|\sigma_z^{j_2}|n\rangle$, etc. 

\noindent This last scheme  differs from \textit{scheme 1} starting from the second division:  it divides both $\mathbb{S}_\pm$  according to $j_2$, where $j_2$ is chosen such that it has the second largest $\langle \tilde{\tau}_z^j\sigma_z^j\rangle$; \textit{Scheme 1} instead treats $\mathbb{S}_\pm$  individually as two instances of a new system with $2^{L-1}$ eigenstates, so $j_M$ in $\mathbb{S}_+$ may differ from $j_M$ in $\mathbb{S}_-$ and $j_M$ is chosen such that it has largest $\langle \tilde{\tau}_z^j\sigma_z^j\rangle$ within the new system.

\textit{Scheme 1}, our chosen algorithm, gives the most local results, especially for smaller disorder $W$ (see Fig.~\ref{fig:AveH_seq}). For the same disorder $W$, \textit{scheme 2} gives less local results for $\tau_z^j$ with larger $j$, because the sector division for large $j$ is constrained by the sectors of small $j$, which is what  motivated us to start with the most local spin. The LIOMs generated using \textit{scheme 3} are almost as local as using \textit{scheme 1}, however with respect to the interactions, \textit{scheme 3} gives larger interaction strength than \textit{scheme 1}, even if it still shows exponential decay with a comparable decay rate. These two observations suggest that \textit{scheme 3} has higher frequency of delocalized events than \textit{scheme 1}. 

Among the three schemes we have shown and many others we have tried, \textit{scheme 1} gives the best result, and we believe it successfully captures the localized cases due to the similarity with \textit{scheme 3}. However, we cannot exclude the possibility that there might be another scheme which gives even lower frequency of delocalized cases.

\begin{figure}[t!]\centering
\includegraphics[width=0.48\textwidth]{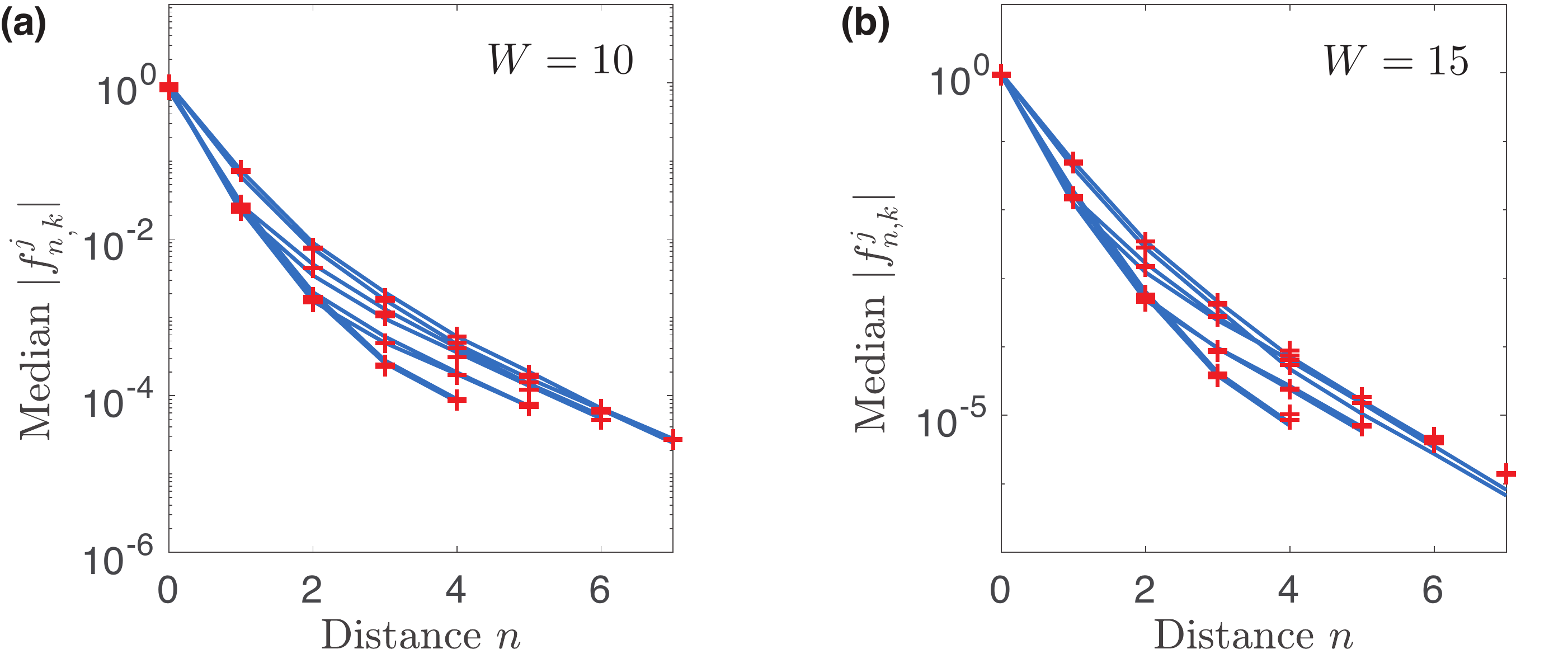}
\caption{\label{fig:lbits1_lbits2}
Median of $|f_{n,k}^j|$ as a function of distance $n$. Red crosses (blue curves) are from $\{\tau_z^j\}$ ($\{\tilde{\tau}_z^j\}$). Different curves represent $j=1,2,\cdots L$. (a) $W=10$. and (b) $W=15$. The two algorithm are almost indistinguishable. System size $L=8$. Median is taken over the index $k$ in $|f_{n,k}^j|$ as well as 100 different disorder realizations.}
\end{figure}

One could also define LIOMs by maximizing their overlap with the corresponding physical spin operators, $\tau_z=\tilde{\tau}_z$, i.e. $a_n^j=-1$ for the $2^{L-1}$ eigenstate $|n\rangle$ with larger $\langle n|\sigma_z^j|n\rangle$ and $a_n^j=1$ for others. By requiring maximum overlap, $\{\tilde{\tau}_z^j\}$ are in principle more local than $\{\tau_z^j\}$ which are described in the main text, but they are not mutually independent and thus cannot form a complete basis. Numerical results for the two constructions are presented in Fig.~\ref{fig:lbits1_lbits2}, showing that the LIOMs generated by the two methods display no significant difference, even for moderate disorder strength $W=10$.

\subsection{Computational complexity}
\begin{figure}[b!]\centering
\includegraphics[width=0.45\textwidth]{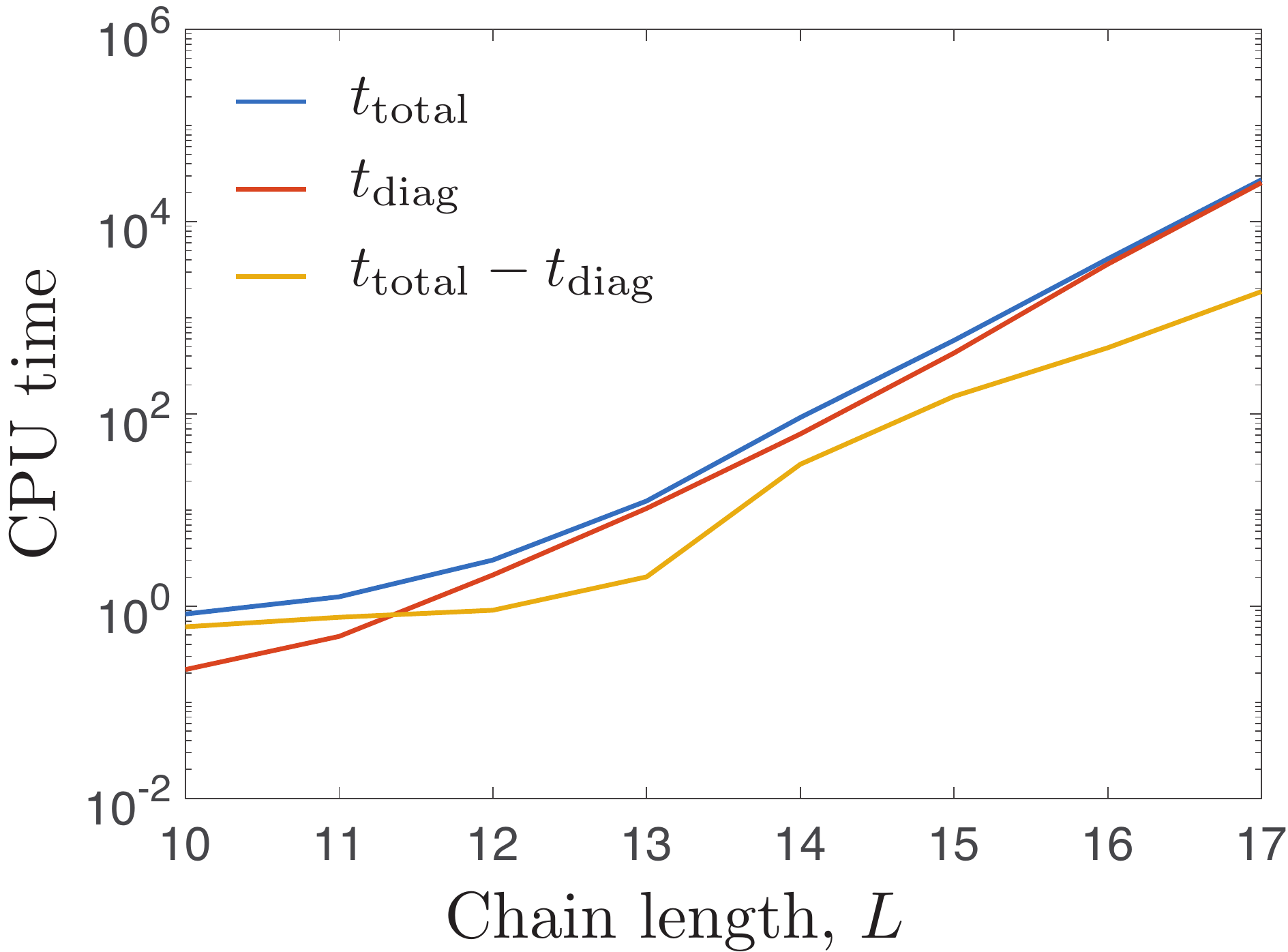}
\caption{\label{fig:cputime}
CPU time for calculating LIOMs under a single disorder realization as a function of the system dimension (spin chain length $L$).
}
\end{figure}
The computational complexity of our LIOMs construction is set by the diagonalization, which is $O(2^{3L})$ (Fig. \ref{fig:cputime}). We note that there are several methods to reduce this complexity and obtain an approximate diagonalization in the localized phase \cite{Khemani2016a,You2016}.

Here we then analyze only the computation complexity of the other steps of the algorithm, which are particular to our scheme:

\begin{itemize}
    \item The complexity of evaluating $\langle n|\sigma_z^j|n\rangle$ for all $j$ and $n$ is $O(L2^{2L})$ because $\sigma_z^j$ is sparse.
    \item The complexity of the recursion step (sorting eigenstates and dividing into sectors) is $O(L^3 2^L)$: for a sector containing $2^N$ states, sorting the $\langle n|\sigma_z^j|n\rangle$ for each $j$ is $N2^N$. For all $j$ is thus $N^2 2^N$. There are $2^{(L-N)}$ such sectors. Total complexity is $\sum_{N=1}^L N^2 2^N 2^{L-N}\sim L^3 2^L$.
    \item Assigning $a_n^k$ has a cost $O(L2^L)$
    \item The decomposition of $H$ to LIOM basis is $O(L2^{L})$, because only diagonal elements are nonzero. 
\end{itemize}
Figure \ref{fig:cputime} confirms that for large $L$, most of the time is spent on diagonalization. 

The only operation that could lead to a complexity higher than $O(2^{3L})$ is computing LIOMs in the physical spin basis, which is required for calculating $f_n^k$. Transferring each LIOM from the eigenbasis to the physical spin basis is $O(2^{3L})$, so total is $O(L2^{3L})$. This high cost is the reason why we studied an alternate metric for the operator localization (the overlap with the physical spin operator), see \ref{app:LiomSpins}.

\end{document}